\documentclass[conference]{IEEEtran}
 \usepackage{upgreek}
 \usepackage{siunitx}
 \usepackage{amssymb,amsmath}
 \usepackage{cite}
 \usepackage{graphicx}
 \usepackage[tight,footnotesize]{subfigure}
 \usepackage{pifont}
 \usepackage{dblfloatfix}
 \usepackage{booktabs}
 \usepackage{hyperref}
\begin{document}
\bstctlcite{IEEEexample:BSTcontrol}
\title{Cryogenic MOSFET Threshold\,Voltage Model}
\author{\IEEEauthorblockN{Arnout Beckers,
		Farzan Jazaeri, and
		Christian Enz}
	\IEEEauthorblockA{Integrated Circuits Laboratory (ICLAB), Ecole Polytechnique F\'ed\'erale de Lausanne, Neuch\^atel, Switzerland\vspace{-0.5cm}}
}
\maketitle
\begin{abstract}
This paper presents a physics-based model for the threshold voltage in bulk MOSFETs valid from room down to cryogenic temperature (4.2\,K). The proposed model is derived from Poisson's equation including bandgap\,widening, intrinsic carrier-density scaling, and incomplete ionization. We demonstrate that accounting for incomplete ionization in the expression of the threshold voltage is critical for an accurate estimation of the current. The model is validated with our experimental results from nMOSFETs of a 28-nm CMOS process. The developed model is a key element for a cryo-CMOS compact model and can serve as a guide to optimize processes for high-performance cryo-computing and ultra-low-power quantum computing. 
\end{abstract}

\section{Introduction}
\vspace{-0.1cm}
Remarkable progress towards a silicon quantum computer in recent years 
has attracted attention to the operation of VLSI technologies at deep-cryogenic temperatures ($<$\,10\,K)~\cite{veldhorst2015two,bourdet,charbon2016cryo,essderc,jeds,solidstate}. Dedicated cryo-CMOS circuits can reduce the overhead associated with the initialization, manipulation, coupling, and read-out of qubits. The power consumption of the CMOS circuits envisioned for qubit control is severely limited ($\approx $ 1\,W at 4.2\,K). At the same time, the circuits need to meet sharp timing constraints on the control pulses applied to the qubits. This delicate trade-off between power consumption and performance at cryogenic temperatures ultimately asks for an accurate model of the MOSFET threshold voltage ($V_T$).

Besides quantum computing, temperature scaling can also provide a performance booster for conventional computing applications, if harnessed carefully. As shown in Fig.\,\ref{fig:meas}, cryogenic operation does \emph{not} improve the on-state current ($I_{ON}$) of small commercial devices at the maximum supply voltage ($V_{DD}$). Furthermore,
the available overdrive voltage ($V_{DD}-V_T$) is reduced by the substantial increase of  $V_T$ down to cryogenic temperatures ($\Delta V_T\approx$~0.1\,\textendash\,0.2\,V). However, there is room for improvement since the leakage current is extremely low at cryogenic temperature. The $I-V$ curve can thus be shifted by selecting a custom work function difference and/or doping, hence supporting a larger $V_{DD}-V_T$ for high performance applications or a scaled $V_{DD}$ for ultra-low power applications. 

Importantly, both technology optimization and the use of commercial devices require that $V_T$ can be accurately predicted. Since the MOSFET turns on over a few millivolts at cryogenic temperatures, a small mis-prediction of $V_T$ leads to orders of magnitude difference in the drain current. To arrive at the most accurate model for $V_T$, incomplete ionization of the dopants and other low-temperature phenomena need to be taken into account by starting from the physics. While many models have been proposed so far for the $V_T$ at low temperature, they are either semi-empirical, validated only down to moderate cryogenic temperature, or not accounting for the various physical phenomena contributing to the temperature dependence of $V_T$~\cite{gutierrez2000low,fox1987mosfet,akturk,dao}. For instance, the physics-based model from Jaeger and Fox~\cite{fox1987mosfet} is valid down to 77\,K and does not account for the dopant freezeout. Recently, Dao et al.\cite{dao} derived a $V_T$ model valid down to 6\,K by introducing two empirical parameters to account for the field-assisted ionization of the dopants. In this work, we demonstrate that both freezeout and field-assisted ionization are accounted for by a single incomplete-ionization term in the Poisson-Boltzmann equation. This results in an entirely physics-based, long-channel $V_T$ model without introducing fitting parameters.
\section{Cryogenic Measurements in 28-nm CMOS}
\begin{figure}[t]
	\centering
	\includegraphics[scale=0.5]{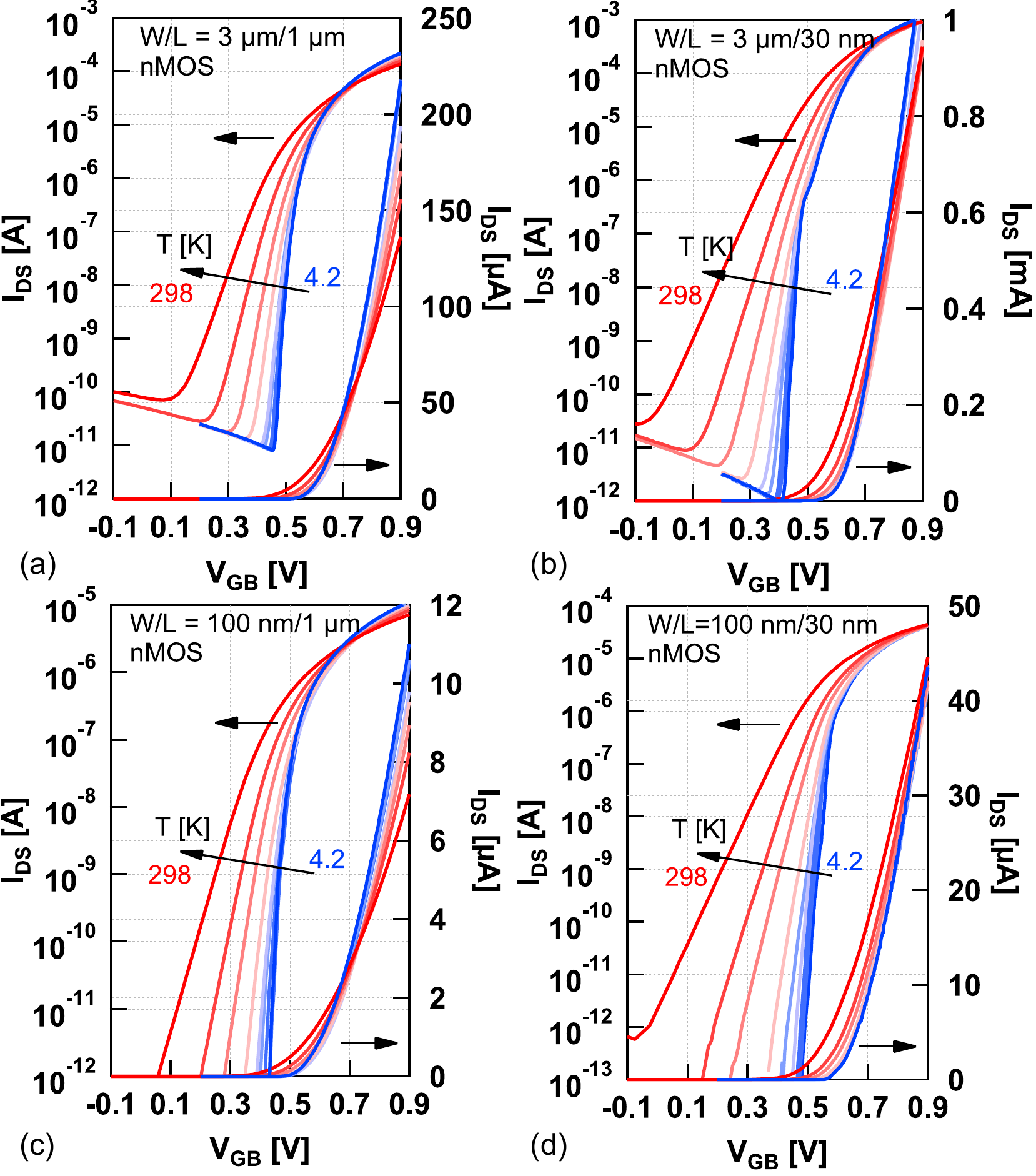}
	\vspace{-0.3cm}
	\caption{Low-temperature measurements down to 4.2\,K in a 28-nm CMOS process ($V_{DS}=$\,0.9\,V). In contrast with large devices (a), small devices (d) do not typically feature a cryogenic temperature-induced improvement of $I_{ON}$ at $V_{DD}=0.9$\,V. Shifting the $I-V$ curve through technology optimization can solve this problem but requires accurate knowledge of $V_T$.} 
	\label{fig:meas}
\end{figure}

\begin{figure}[t]
	\centering
	\includegraphics[scale=0.53]{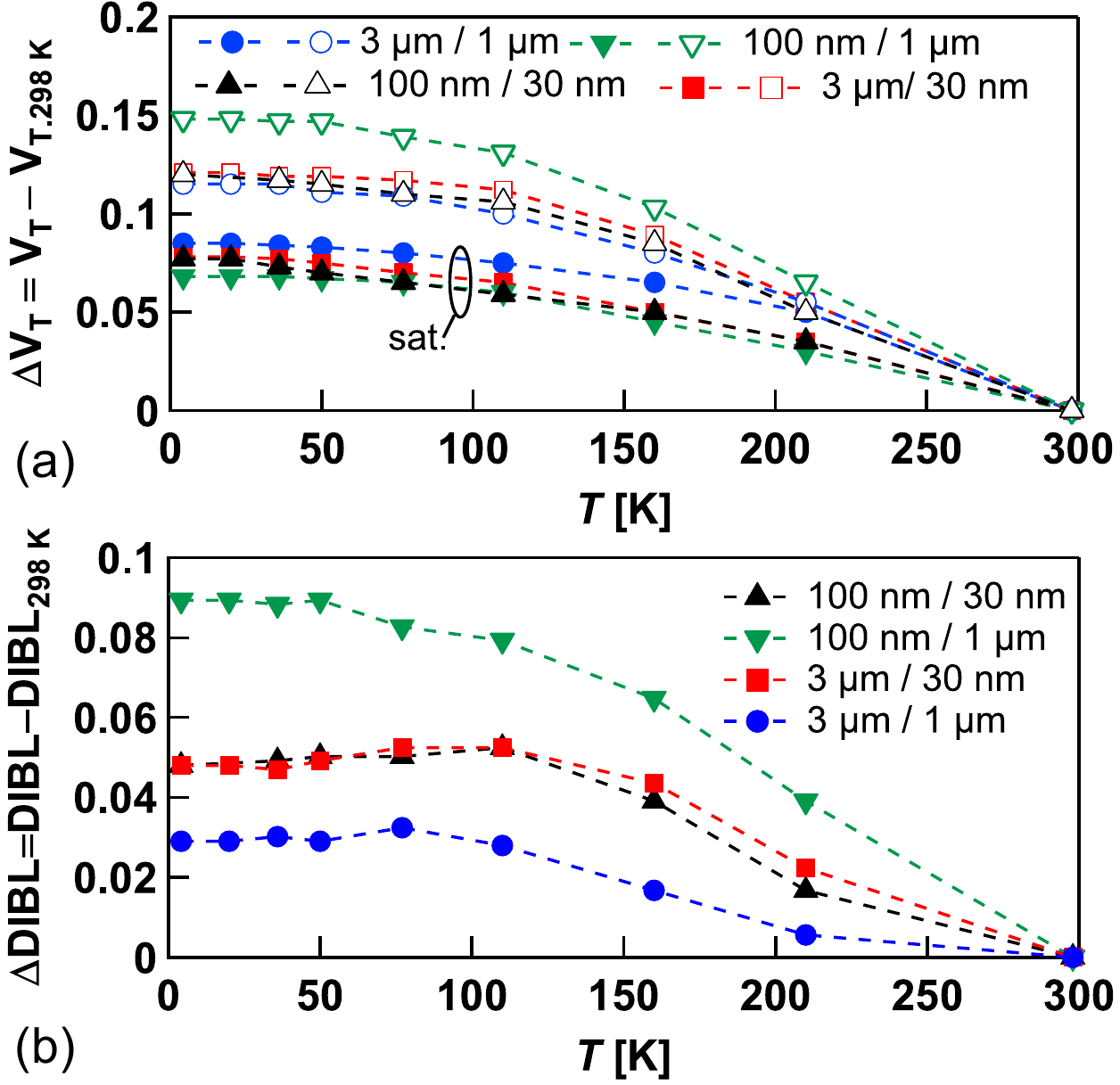}
	\vspace{-0.3cm}
	\caption{a) Shift in $V_T$ in saturation from Fig.\,\ref{fig:meas} ($V_{DS}=$\,0.9\,V, filled markers) and linear ($V_{DS}=\,$5\,mV, open markers). b) DIBL versus temperature.}
	\label{fig:dibl}
\end{figure}

Low-temperature measurements were performed in the four $n$-type width-over-length device-corners of a 28-nm bulk CMOS process. After cooling a Lakeshore CPX cryogenic probe station to its 4.2-K base temperature, the station was heated to 298\,K with intermediate temperature steps at 20, 36, 50, 77, 110, 160, and 210\,K. The transfer characteristics in linear ($V_{DS}$\,=\,5\,mV) and saturation ($V_{DS}$\,=\,0.9\,V) were acquired using a Keysight B1500A device analyzer. Temperature-variable probes were used on the bulk and source contact pads to ensure the same bulk reference at the different temperature steps. The measurements in saturation are shown in Fig.\,\ref{fig:meas}. $V_T$ was extracted using a linear extrapolation method from the maximum transconductance and $\Delta V_T$ is shown in Fig.\,\ref{fig:dibl}(a). For all devices, a monotonic $V_T$ increase is obtained, leveling off in the deep cryogenic regime (below $\approx$\,50\,K). The drain-induced barrier lowering (DIBL) [Fig.\,\ref{fig:dibl}(b)] follows a similar trend and is highest for a narrow-long device. 

\section{Overview of $V_T$ Temperature Dependences}
We assume an $n$-type MOSFET with uniform acceptor doping density ($N_A$) in the silicon body [Fig.\,\ref{fig:banddiagram}(a)]. The source and drain junctions can be assumed completely ionized at all temperatures due to impurity-band formation as a result of degenerate doping~\cite{akturk}. Based on the MOSFET band diagrams in Fig.\,\ref{fig:banddiagram}, we discuss the physical phenomena that impact the $V_T$ of a large MOSFET.  The vacuum energy level ($E_{vac}$) is taken as the reference and a potential drop ($\psi$) is defined with respect to the intrinsic energy level $E_i$ in the bulk ($E_{i,b}$), i.e.,  $\psi=-(E_i-E_{i,b})/q$, where $q$ is the elementary charge. 
\subsubsection{Bandgap Widening and Scaling of Fermi-Dirac Occupation Function}
\begin{figure*}[t]
	\flushleft
	\includegraphics[width=\textwidth]{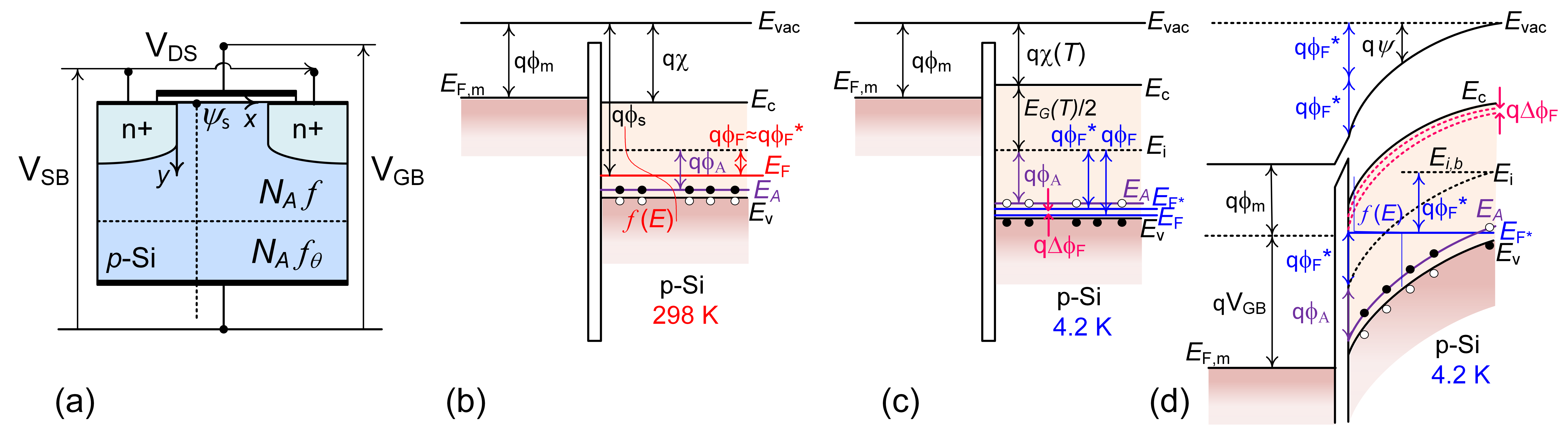}
	\vspace{-0.8cm}
	\caption{a) Cross-section of an $n$MOSFET with incomplete ionization of $N_A$ (thermal ionization probability given by $f_\theta$. and thermal + field-assisted ionization given by $f=\,$Fermi-Dirac). Band diagrams from interface to bulk are shown: b) flat-band at 298\,K, c) flat-band at 4.2\,K ($p_p=N_A^-$ gives $\Phi_\mathrm{F}^*$, $p_p=N_A$ gives $\Phi_\mathrm{F}$), and d) strong-inversion threshold in the strict meaning ($n_p=N_A^-$). To reach this point, a band bending of $2q\Phi_\mathrm{F}^*$ is needed when including incomplete ionization. An additional band bending over $q\Delta \Phi_\mathrm{F}=q\Phi_\mathrm{F}-q\Phi_\mathrm{F}^*$ is needed to increase the inverted electron density from $n_p=N_A^-$ to $n_p=N_A$.}
	\label{fig:banddiagram}
\end{figure*}

The Si bandgap increases from $\approx$~1.12 to 1.17 eV from 298 to 4.2\,K, plotted in Fig.\,\ref{fig:phif}. The Fermi-Dirac occupation function $f(E)$ scales exponentially, approaching a step function at 4.2\,K [compare $f(E)$ in Fig.\,\ref{fig:banddiagram}(b) and (d)]. More band bending is required to generate sufficient overlap of $f(E)$ with the density-of-states ($DOS$) in the conduction band. Both phenomena will thus act to increase $V_T$. The effective masses and the $DOS$ are assumed to have a negligible temperature dependence compared to $f(E)$\cite{gutierrez2000low}. 
\subsubsection{Intrinsic Carrier-Density Scaling}
The Boltzmann statistics is validated for the deep-cryogenic temperatures~\cite{tedpaper}. The intrinsic carrier density ($n_i$) is thus given by the expression with exponential dependence on $-E_g$ and $1/T$. Respectively, the $E_g$ widening and $f(E)$ scaling thus strongly decrease $n_i$.
é
\subsubsection{Increase in Bulk Fermi-potential}
The decrease in $n_i$ is sufficiently strong to make the Fermi potential (assuming complete ionization) increase, given by $\Phi_\mathrm{F}=U_T\ln(N_A/n_i)$ [Fig.\,\ref{fig:banddiagram}(b) vs. \ref{fig:banddiagram}(c)], despite its direct dependence on $U_T=kT/q$, the thermal voltage. 
Indeed, in order to satisfy charge neutrality in the case of a completely ionized $N_A$ at 4.2\,K [$p_p=n_i\exp(\Phi_\mathrm{F}/U_T)=N_A$], $E_F$ needs to lie very close to the valence band edge ($E_v$) to compensate for $n_i$ [Fig.\ref{fig:banddiagram}(c)]. The increase in $\Phi_\mathrm{F}$ ranges from about 0.1\,V to 0.5\,V depending on $N_A$, as shown by the dashed lines in Fig.\,\ref{fig:phif}. It can be checked that $\lim_{T\rightarrow 0\,\mathrm{K}}\Phi_\mathrm{F}=E_g/2$. This increase in $\Phi_\mathrm{F}$ constitutes a large portion of the $V_T$ increase. However, the $\Phi_\mathrm{F}$ increase is too strong and not sufficiently accurate. At 4.2\,K, the silicon body would be degenerate ($E_g/2-\Phi_\mathrm{F}<3kT$), which does not stroke with the measured electrical behavior in MOSFETs at 4.2\,K. Incomplete ionization of $N_A$ needs to be taken into account in the Fermi potential to avoid this. 
\subsubsection{Incomplete Ionization (thermal)}
\begin{figure}[t]
	\centering
	\includegraphics[scale=0.5]{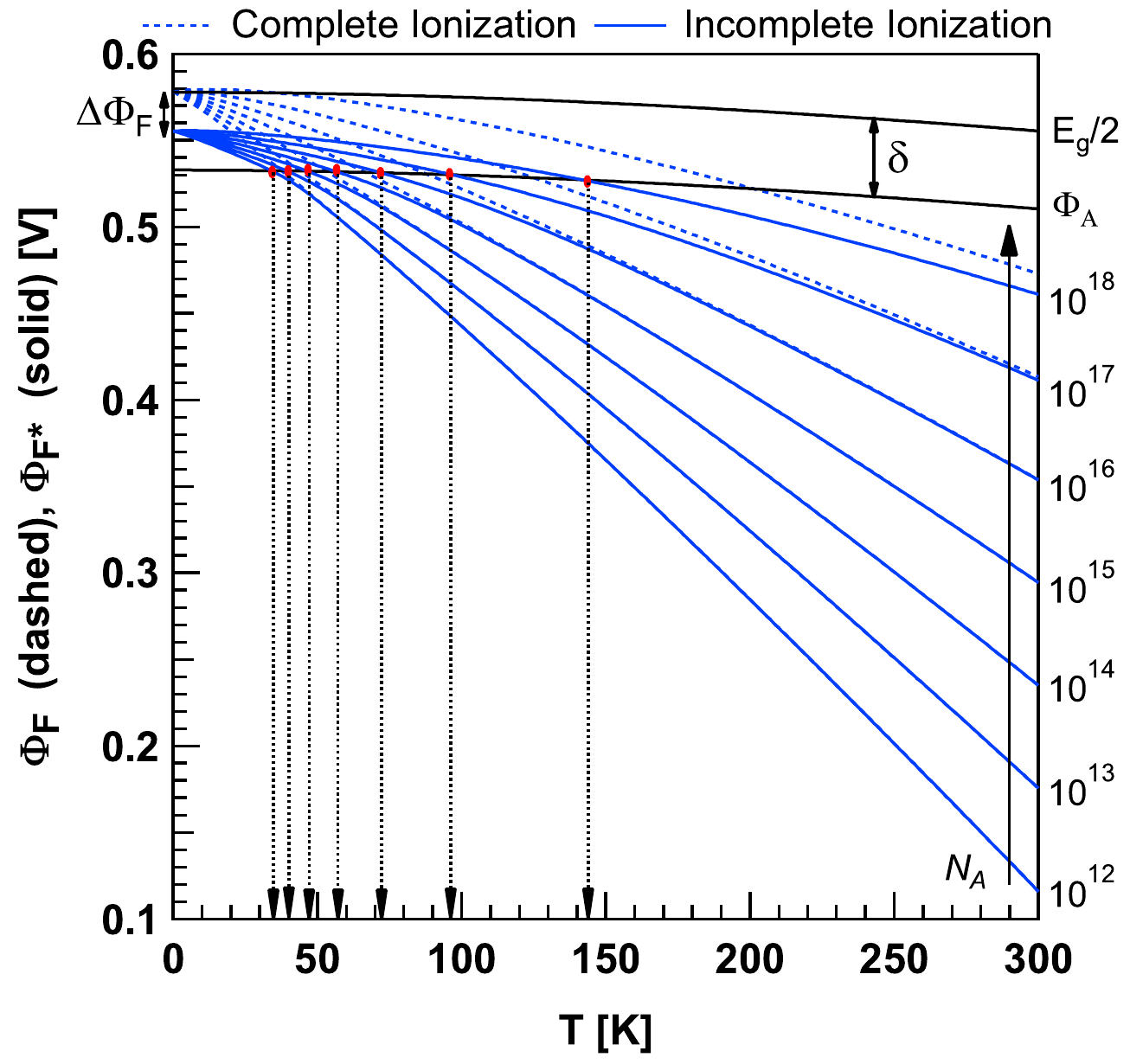}
	\vspace{-0.3cm}
	\caption{Fermi potential in an $n$MOSFET as a function of $T$ and $N_A$  assuming complete ionization $\left[\Phi_\mathrm{F}=U_T\ln(N_A/n_i)\right]$ versus incomplete ionization $\left[\Phi_\mathrm{F}^*\text{ from (\ref{phif})}\right]$. Freezeout of the $p$-type bulk happens when $\Phi_\mathrm{F}^* \geqslant \Phi_\mathrm{A} $ [= when $E_F^*\leqslant E_A$ in Fig.\,\ref{fig:banddiagram}(c)]. Critical freezeout temperatures can be identified for given $N_A$ at $\Phi_{\mathrm{F}}^*=\Phi_\mathrm{A}$. The difference between $\Phi_\mathrm{F}$ and $\Phi_\mathrm{F}^*$ is $\Delta \Phi_\mathrm{F}$.}
	\label{fig:phif}
\end{figure}

\begin{figure}[t]
	\centering
	\includegraphics[width=0.38\textwidth]{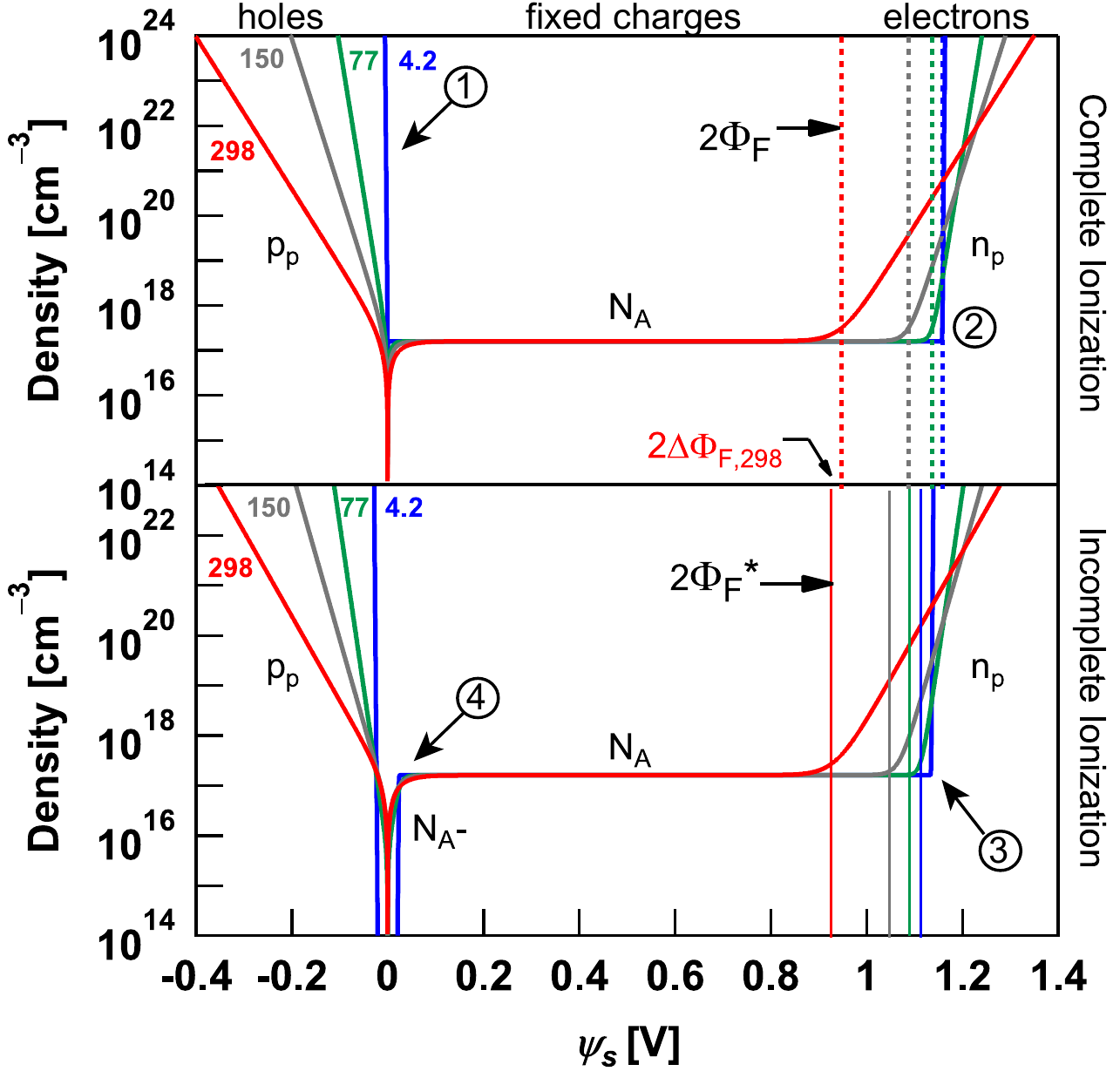}
	\vspace{-0.3cm}
	\caption{Surface charge densities. (1) $f(E)$ scaling leads to a steeper turn-on of $p_p$ in accumulation, and, similarly, of $n_p$ in inversion. (2) Both $2\Phi_\mathrm{F}$ and $2\Phi^*_\mathrm{F}$ increase when temperature decreases. (3) The lower $2\Phi^*_\mathrm{F}$ vs. $2\Phi_\mathrm{F}$ is crucial at 4.2\,K leading to a much lower inverted electron density. (4) Delayed turn-on of the fixed charges at 4.2\,K due to the gate-voltage-dependent ionization of the frozen-out dopants. ($N_A=\SI{e18}{\per\cubic\centi\meter}$, $V_{ch}=0$).}
	\label{fig:rho}
\end{figure}

Returning back to the charge neutrality, this gives $p_p= N_A^-=N_Af(E_A)$, where $f(E_A)$ is the ionization probability of the dopants, both due to temperature (thermal ionization) and the applied voltages (field-assisted ionization). Under charge neutrality, the field is zero and thus $f(E_A)$ equals the thermal ionization probability, given by $f_\theta=1/\{1+\alpha\exp\left[\Phi_{\mathrm{F}}^*+V_{SB}\right]\}$, plotted in Fig.\,\ref{fig:ftheta}(b). Below $f_\theta$ $\approx$ 20\%, the silicon body will enter freezeout. At 4.2\,K, incomplete ionization results in the position of $E_F^*$ instead of $E_F$ in Fig.\,\ref{fig:banddiagram}(c), or mathematically: 
\begin{equation}
\Phi_\mathrm{F}^*=\underbrace{U_T\ln\frac{N_A}{n_i}}_{\Phi_\mathrm{F}}-\underbrace{U_T\ln\frac{1+\sqrt{1+(4\alpha N_A)/n_i}}{2}}_{\Delta \Phi_\mathrm{F}}, 
\label{phif}
\end{equation}
where $\alpha=g_A\exp{[(E_A-E_i)/(kT)]}=g_A\exp{(-\Phi_A/U_T)}$. The acceptor potential $\Phi_A\triangleq (E_{i}-E_{A})/q$ as well as $\Phi_\mathrm{F}^*$ are shown in Fig.\,\ref{fig:phif} with solid lines. The function $\Delta \Phi_\mathrm{F}=\Phi_\mathrm{F}-\Phi_\mathrm{F}^*$ is plotted in Fig.\,\ref{fig:ftheta}(a) for different $N_A$ and is in the order of millivolts. It can be checked that $\lim_{T\to 0\,\mathrm{K}}\Delta \Phi_\mathrm{F}=\delta/2$  with $\delta=E_A-E_v=0.045\,\si{\electronvolt}$, for typical, hydrogen-like Si:B doping and thus $\lim_{T\to 0\si{K}}\Phi_\mathrm{F}^*=\Phi_A+\delta/2$ as shown in Fig.\,\ref{fig:phif}. It might seem that $\Delta \Phi_\mathrm{F}$ due to incomplete ionization is negligible for $V_T$ at cryogenic temperatures compared to the overall increase of $\Phi_\mathrm{F}$ or $\Phi_\mathrm{F}^*$ in Fig.\,\ref{fig:phif}. However, Fig.\,\,\ref{fig:rho} shows a large discrepancy in inverted density ($n_p$) at $2\Phi_\mathrm{F}$ and $2\Phi_\mathrm{F}^*$. 

\subsubsection{Incomplete Ionization (field)}
A second effect associated with the incomplete dopant ionization is the process of the frozen-out dopants ionizing due to the field, as shown by the band bending in Fig.\,\ref{fig:banddiagram}(d). However, this field-assisted process of ionization (e.g., Poole-Frenkel~\cite{foty}) happens in early depletion and is already completed near the surface before reaching the on-set of strong inversion where $V_T$ is evaluated. This process can thus be possibly neglected in $V_T$ and is not a justification for fitting parameters as done in\cite{dao}. 
  
\section{\label{sec:model}Analytical $V_T$ Model} 
From standard MOS device physics it follows that the gate-to-bulk voltage is given by $V_{GB}=\psi_s+\Phi_{ms}+\varepsilon_{si}\mathcal{E}_s(\psi_s)/C_{ox}$, where $\psi_s$ is the surface potential, $\varepsilon_{si}$ the silicon permittivity, $\mathcal{E}_s$ the electric field normal to the semiconductor/oxide interface, $C_{ox}$ the gate oxide capacitance, and $\Phi_{ms}$ the difference between the metal ($\Phi_m$) and the semiconductor ($\Phi_s$) work functions. Note that we do not take into account the oxide and interface-trapped charges. The threshold voltage is usually defined as the $V_{GB}$ evaluated at a given value of $\psi_s$, i.e., the inversion threshold $\psi_s'$, for which $n_p$ equals the \emph{totally ionized} $N_A$. Note that in the strict meaning of the inversion threshold when incomplete ionization is present, we have to define the threshold at $n_p=N_A^-$, since then the density of carriers in the channel has been inverted from the flatband case ($p_p=N_A^-$). However, this density can be very low due to freezeout and it is thus more meaningful to keep defining the threshold at $n_p=N_A$.
The temperature dependence of the four terms of $V_T$ are investigated separately below: 
\subsubsection{\label{sec:invthresh}$\psi_s'$} Near 298\,K, the inversion threshold is taken at $2\Phi_\mathrm{F}$ since then $n_p=N_A$ when complete ionization can be assumed. However, this threshold needs to be reassessed for the low and cryogenic temperatures. From $n_p=N_A$, we find that $\psi_s'=U_T\ln(N_A/n_i)+\Phi_\mathrm{F}^*+V_{ch}(x)+V_{SB}$, where $V_{ch}$ is the channel voltage. We evaluate $V_T$ at source ($x=0$) and assume $V_{SB}=0$. It can be checked that this yields back $2\Phi_\mathrm{F}$ for complete ionization ($\alpha=0$). However, using (\ref{phif}), the correct inversion threshold is $\psi_s'=\Phi_\mathrm{F}+\Phi_\mathrm{F}^*=2\Phi_\mathrm{F}^*+\Delta \Phi_\mathrm{F}=2\Phi_\mathrm{F}-\Delta \Phi_\mathrm{F}$, and not 2$\Phi_\mathrm{F}$,\,nor 2$\Phi_\mathrm{F}^*$.
\subsubsection{$\Phi_{ms}$} From Fig.\,\ref{fig:banddiagram} we can obtain that $\Phi_{ms}\,=\,\Phi_m-\left(\chi+E_g/2+\Phi_\mathrm{F}^*\right)$. $E_{F,m}$ is assumed temperature independent due to metallic gate (or degeneracy in poly-Si). Due to lack of available data on the temperature dependence of $\chi$ in silicon down to 4.2\,K, it is assumed that $\chi$ decreases with the same amount as $E_g$ increases. Under these assumptions, the temperature dependence of $\Phi_{ms}$ is defined by $\Phi_\mathrm{F}^*$. 
\subsubsection{$\varepsilon_{si}\mathcal{E}_s(\psi_s')/C_{ox}$} 
\begin{figure}[t]
	\centering
	\includegraphics[scale=0.55]{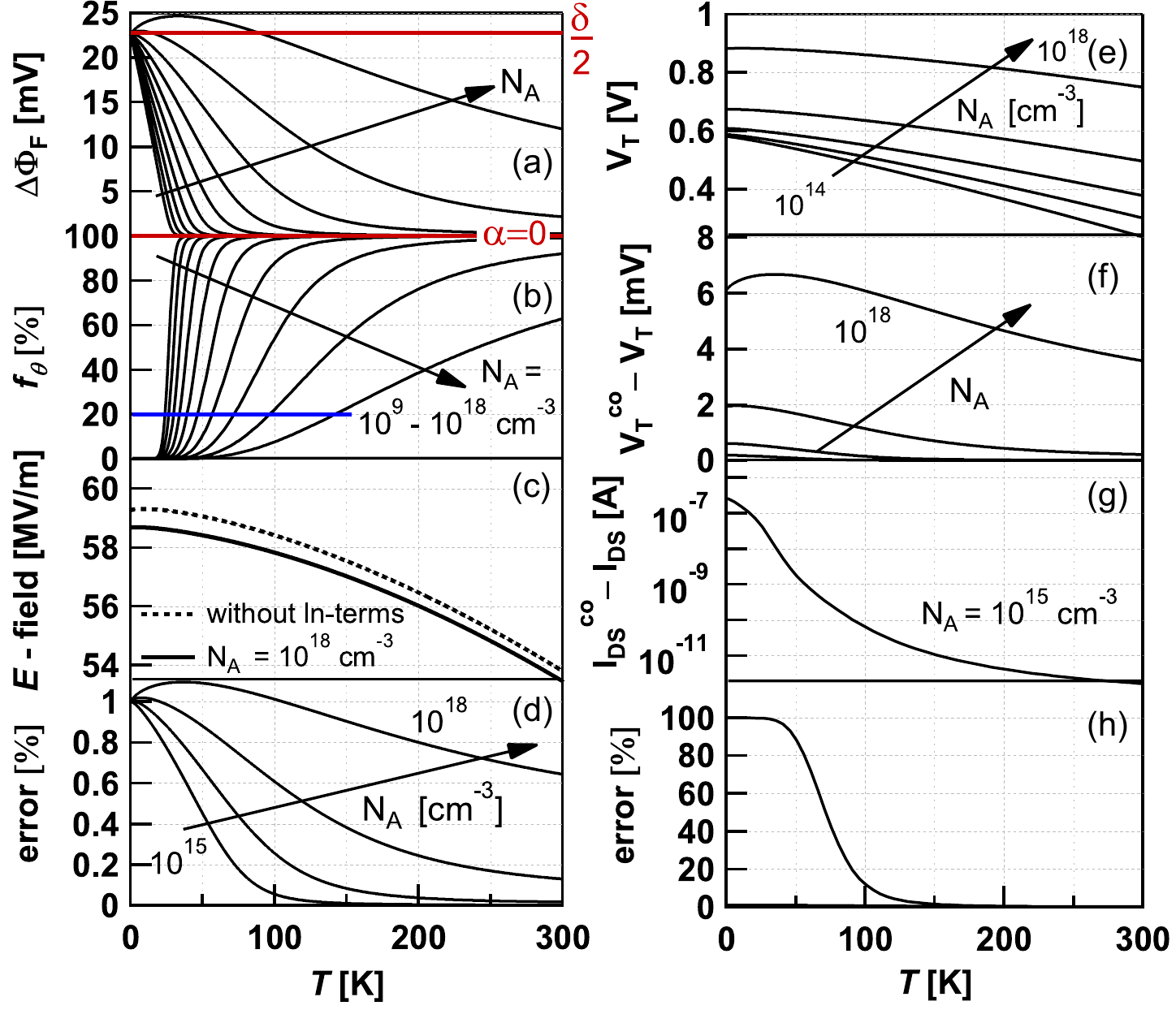}
	\vspace{-0.3cm}
	\caption{a) $\Delta \Phi_\mathrm{F}=\Phi_\mathrm{F}-\Phi_\mathrm{F}^*$, b) Thermal ionization probability of the dopants in flat-band and $V_{SB}=0$. Complete ionization for $\alpha=0$ and freezeout for $f_\theta \leqslant $\,20\,\%. c)-d) $\mathcal{E}_s$ approximation without ln-terms. e) $V_T$ from (\ref{eq:VT}), f) Absolute error $V_T^{co}-V_T$, g)-h) Absolute and relative error on $I_{DS}$.}
	\label{fig:ftheta}
\end{figure}

\begin{figure}
	\centering
	\includegraphics[scale=0.58]{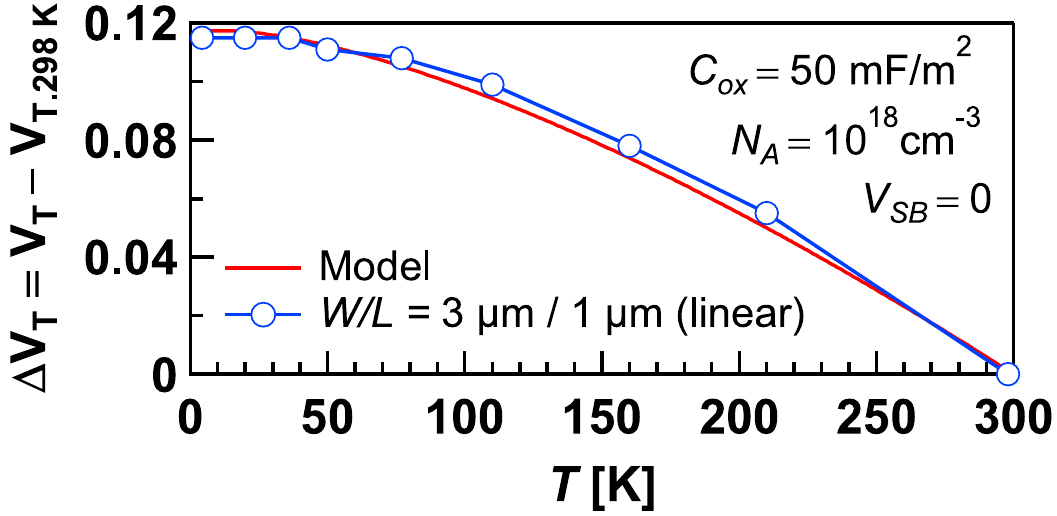}
	\vspace{-0.3cm}
	\caption{Model validation with experimental results in large $n$MOS.}
	\label{fig:validation}
\end{figure}

The temperature dependences of $\varepsilon_{si}$ and $C_{ox}$ can be assumed negligible. From the Poisson-Boltzmann equation, $\partial^2\psi(y)/\partial y^2=-\rho(\psi)/\varepsilon_{si}$, including incomplete ionization, $\mathcal{E}_s$ in depletion can be derived as~\cite{tedpaper}: $\mathcal{E}_s=(U_T/L_D)\sqrt{\psi_s/U_T+\mathcal{F}}$, where
$\psi_s$ is the surface potential, $L_D=\sqrt{\varepsilon_{si}U_T/(2qN_A)}$, and $\mathcal{F}$ an extra term due to the incomplete ionization of $N_A$, given by $\mathcal{F}=\ln f_\theta+\ln\{1+\alpha\exp \left[\left(-\psi_s+\Phi_\mathrm{F}^*+V_{ch}+V_{SB}\right)/U_T\right]\}$. The second term, dependent on $\psi_s$, is due to the field-assisted ionization process. Evaluating $\mathcal{E}_s$ at $\psi_s'=2\Phi_\mathrm{F}-\Delta \Phi_\mathrm{F}+V_{ch}+V_{SB}$ gives 
\begin{equation}
\mathcal{E}_s(\psi_s')=\frac{U_T}{L_D}\sqrt{\frac{2\Phi_{\mathrm{F}}-\Delta \Phi_\mathrm{F}}{U_T}+\ln f_\theta+\ln\left(1+\alpha e^{\frac{-\Phi_\mathrm{F}}{U_T}}\right)}, 
\label{es}
\end{equation} 
for $V_{ch}$\,$=$\,$V_{SB}$\,$=$\,0. As evidenced by Figs.\,\ref{fig:ftheta}(c)-(d), the two logarithmic terms can be neglected tolerating a $\approx$\,1\% error on the electric field. This approximation can be viewed as neglecting the field-assisted ionization process which is completed at threshold ($\psi_s\gg0$). Note that the different inversion threshold due to incomplete ionization, $2\Phi_\mathrm{F}-\Delta \Phi_\mathrm{F}$, is still taken into account. Considering all the temperature dependences together from the previous subsections, $V_T$ is derived at the source as: 
\begin{equation}
V_T=2\Phi_\mathrm{F}-\Delta \Phi_\mathrm{F}+V_{SB}+\Phi_m'-\Phi_\mathrm{F}^*+\Gamma_b\sqrt{2\Phi_{\mathrm{F}}-\Delta \Phi_\mathrm{F}+V_{SB}}. 
\label{eq:VT}
\end{equation} 
where $\Phi_m'=\Phi_m-(\chi+E_g/2)$ and $\Gamma_b=\sqrt{2qN_A\varepsilon_{si}}/C_{ox}$. It is now interesting to develop the error introduced by assuming complete ionization at all temperatures in (\ref{eq:VT}) ($\Phi_\mathrm{F}^*=\Phi_\mathrm{F}$, $\Delta \Phi_\mathrm{F}=0$), reducing $V_T$  to $V_T^{co}=V_T(\psi_s'=2\Phi_{\mathrm{F}})$. The absolute error $V_T^{co}-V_T$ (note that $V_T^{co} > V_T$) is not large (millivolts) compared to a typical $V_T$ at 4.2\,K in a commercial device ($\approx$\,0.4\,V). However, due to the steepness of the subthreshold swing at 4.2\,K, the absolute and relative error on the current are very large at cryogenic temperatures, as shown in Figs.\,\ref{fig:ftheta}(g) and \ref{fig:ftheta}(h). The current is considered in the linear regime, $I_{DS}=-(W/L)\mu Q_m V_{DS}$, and at constant mobility, $\mu$, to illustrate this discrepancy. The used $Q_m$ including incomplete ionization is given by 
\begin{equation}
Q_m=-\gamma\sqrt{\frac{n_i}{N_A}\exp(\beta)+\frac{\psi_s}{U_T}+\mathcal{F}}
+\gamma\sqrt{\frac{\psi_s}{U_T}+\mathcal{F}}, 
\label{qmpoisson}
\end{equation} 
where $\gamma=\varepsilon_{si}U_T/L_D$ and $\beta=(\psi_s-\Phi_{\mathrm{F}}^*-V_{ch}-V_{SB})/U_T$.  Fig.\,\ref{fig:validation} validates the $V_T$ model from (\ref{eq:VT}) with the experimental results in a large $n$MOS at $V_{DS}=$\,5\,mV with negligible DIBL. 
\section{Conclusion}
The increase of $V_T$ down to 4.2\,K is physically explained by $E_g$ widening and Fermi-Dirac scaling. This leads to a strong decrease in $n_i$ and thus increase in the inversion threshold $2\Phi_\mathrm{F}$. Bulk incomplete ionization or freezeout makes the threshold $2\Phi_\mathrm{F}-\Delta \Phi_\mathrm{F}$, where $\Delta \Phi_\mathrm{F}$ is temperature- and doping-dependent. Although in mV-range, $\Delta \Phi_\mathrm{F}$ is significant for the cryogenic current. The process of field-assisted dopant ionization is negligible at inversion threshold. Good agreement is obtained between the $V_T$ model and the experimental results.\vspace{-0.2cm}
\section*{Acknowledgment}\vspace{-0.2cm}
This project has received funding from the European Union's Horizon 2020 Research \& Innovation Programme under grant agreement No. 688539 MOS-Quito. The authors thank AQUA (EPFL) for access to a cryo-probe station, and Dr. Grabinski for guidance with measurement automation.\vspace{-0.2cm}

\bibliographystyle{IEEEtran}

\bibliography{threshold}

\end{document}